\documentclass[10pt,a4paper,twocolumn,conference]{IEEEtran}
\usepackage[margin=1.9 cm]{geometry}
\usepackage{cite}	
\usepackage{graphicx} 
\usepackage[latin1]{inputenc} 
\usepackage[T1]{fontenc} 
\usepackage{amsmath,amsfonts,amsbsy,amssymb} 
\usepackage{mathabx} 
\usepackage{amssymb} 
\usepackage{amsmath}
\usepackage{amsthm}	
\usepackage{mathrsfs}
\usepackage[nolist]{acronym} 
\usepackage{tabularx} 
\usepackage{multirow}
\usepackage{wasysym}
\usepackage{float}
\usepackage{color} 
\usepackage{graphicx}
\usepackage[caption=false]{subfig}
\usepackage{lipsum} 
\usepackage{enumitem} 
\usepackage[font=small,skip=0pt]{caption}
\DeclareCaptionLabelFormat{mylabel}{#1#2.\hspace{1ex}}
\newtheorem{proposition}{Proposition}
\newtheorem{lemma}{Lemma}

\IEEEpubidadjcol 
\IEEEoverridecommandlockouts
\IEEEpubid{\makebox[\columnwidth]{978--1--5386--3873--6/17/\$31.00~\copyright~2017 IEEE\hfill} \hspace{\columnsep}\makebox[\columnwidth]{ }}
 
\begin{document}

\title{On the Performance of Ultra-Reliable Decode and Forward Relaying Under the Finite Blocklength }
\author{Parisa Nouri\IEEEauthorrefmark{0}, Hirley Alves\IEEEauthorrefmark{0}, and Matti Latva-aho \IEEEauthorrefmark{0}\\
	\IEEEauthorblockA{
		\IEEEauthorrefmark{0}Centre for Wireless Communications (CWC), University of Oulu, Finland\\
	}
	%
	%

		firstname.lastname@oulu.fi
		
}

%

\maketitle

\begin{abstract}
In this paper, we examine the performance of the decode-and-forward (DF) relaying protocols with  finite blocklength (FB). We provide the overall outage probability of three distinct DF relaying protocols, where the channels are assumed to be quasi static Rayleigh fading. More importantly, we derive the closed form expressions of the outage probability in the three relaying scenarios. We illustrate protocols where the cooperative communications outperform the direct transmission (DT). In addition, we compare the operating efficiency of the cooperative schemes in the ultra-reliable (UR) region.\\ 
\end{abstract}


\section{Introduction}

The fifth generation (5G) of the cellular systems is expected to deal with the demanding requirements such as latency and reliability in addition to attaining higher data rates in comparison to the previous generations. 5G envisages to handle both the ultra-reliable communication (URC) and the massive machine-to-machine communications. Wireless communication systems need to be designed in such a way to meet the stringent ultra-high reliability and low latency requirements~\cite{7529226}. In massive machine-to-machine communications, a huge number of devices/sensors communicate with short data packets in order to support the low latency necessity in the exigent situations, in a certain domain, e.g. in the smart grid or sensor networks~\cite{7041045}. In the case of URC, short data packets are transfered under the stringent reliability and latency requirements in order to support the connectivity for the total duration of the communication. URC can be applied to different wireless applications, e.g. cloud connectivity, industrial control and safe interconnection between vehicles~\cite{7529226}.

Since in the short packet transmissions, the blocklength of the metadata is analogous to the size of the information bits, a considerable loss occurs as a result of an inefficient encoding of the control information bits~\cite{7529226}. Therefore, in order to cope with the reliability and latency requirements with short messages, we resort to studying the finite-blocklength regime. There are several works which have studied different aspects of FB coding. For instance, authors in~\cite{5452208}, provided a tight approximation of achievable coding rate for a specified outage probability under the short packet transmissions since majority of the theoretical results assumed infinite blocklength codes. Moreover, Makki~\cite{7249552}, studied the performance of the spectrum sharing networks under the FB. They showed that the system efficiency highly depends on the packet length and optimal power allocation strategy boosts the performance of the communication network under the finite-blocklength regime. 

Relaying is a well known technique in order to enhance the reliability~\cite{hu2015performance}. Authors in~\cite{swamy2015cooperative}, propose a method that meet the high reliability and latency requirements via cooperative relaying. Relaying protocol is characterized as fixed, adaptive and feedback protocols. In the fixed protocol, the relay always collaborates with the source and forwards the data packets while in the adaptive protocol, the defined threshold rule enables it to be able to make the decisions if it should retransmit the packet or not independently. In the feedback protocol, the relay cooperates only if the destination asks for the retransmission~\cite{zimmermann2005performance}. Therefore, it would be interesting to exploit it in the context of FB. 

In early works, authors in~\cite{hu2015performance} and~\cite{hu2016blocklength}, studied the relaying performance of the quasi-static Rayleigh channels under the FB regime. 
Authors in~\cite{hu2015performance}, considered a relaying protocol where the channel gains of the direct link and relaying link are accumulated at the receiver under the perfect CSI assumption. They illustrated that the performance depends on the outage probability of the relaying phase. They realized that the performance loss increases when the error probability of the source-to-relay link is higher than the overall error probability. Authors in~\cite{hu2016blocklength}, focused on the effective capacity and the throughput of the relaying protocol with the moderate CSI at the source which means that the average channel gain is higher than the instantaneous channel gain during a certain transmission interval. They showed that the effective capacity is inversely proportional to the blocklength while the throughput smoothly increases in the blocklength and relaying becomes even more advantageous at high signal-to-noise-ratio (SNR) regime. 

However these recent works, concentrated on the relaying scenario without exploiting the ultra-high reliability in the context of FB. Hence, following the concept of the relaying, this paper guarantees the URC under the FB and compare three cooperative relaying protocols, namely, decode-and-forward (DF), selection combining (SC) and maximum ratio combining (MRC). Throughout this paper, we provide a tight approximated overall outage probability and also the closed form expression of the outage probability for each of the relaying scenarios. Moreover, we illustrate that MRC relaying scenario outperforms the other relaying protocols in the blocklength and is more suitable for the URC communications in terms of power consumption, and different from~\cite{hu2015performance} we propose close-form expression for the outage probability. 

{\textbf{Notation}:} Let $f_{W}(\cdot)$ and $F_{W}(\cdot)$ denote the probability density function (PDF) and cumulative distribution function (CDF) of a random variable  (RV) W, respectively. $\operatorname{Q}$-function is defined as  $\operatorname{Q(w)} \!=\! \int_{w}^{\infty} \tfrac{1}{\sqrt{2\pi}}\operatorname{e}^{-t^2/2}dt$~\cite[Eq. F.2]{miller2012probability}, and its inverse is denoted as $\operatorname{Q}^{-1}(\cdot)$. The packet error probability is denoted by  $\epsilon$ and $\operatorname{E[\cdot]}$ is the expectation.

\section{System Model} \label{sc:system_model}
\begin{figure}[!b]
	\centering
	\vspace{0.5cm}
	\includegraphics[width=\columnwidth,height=1.7in]{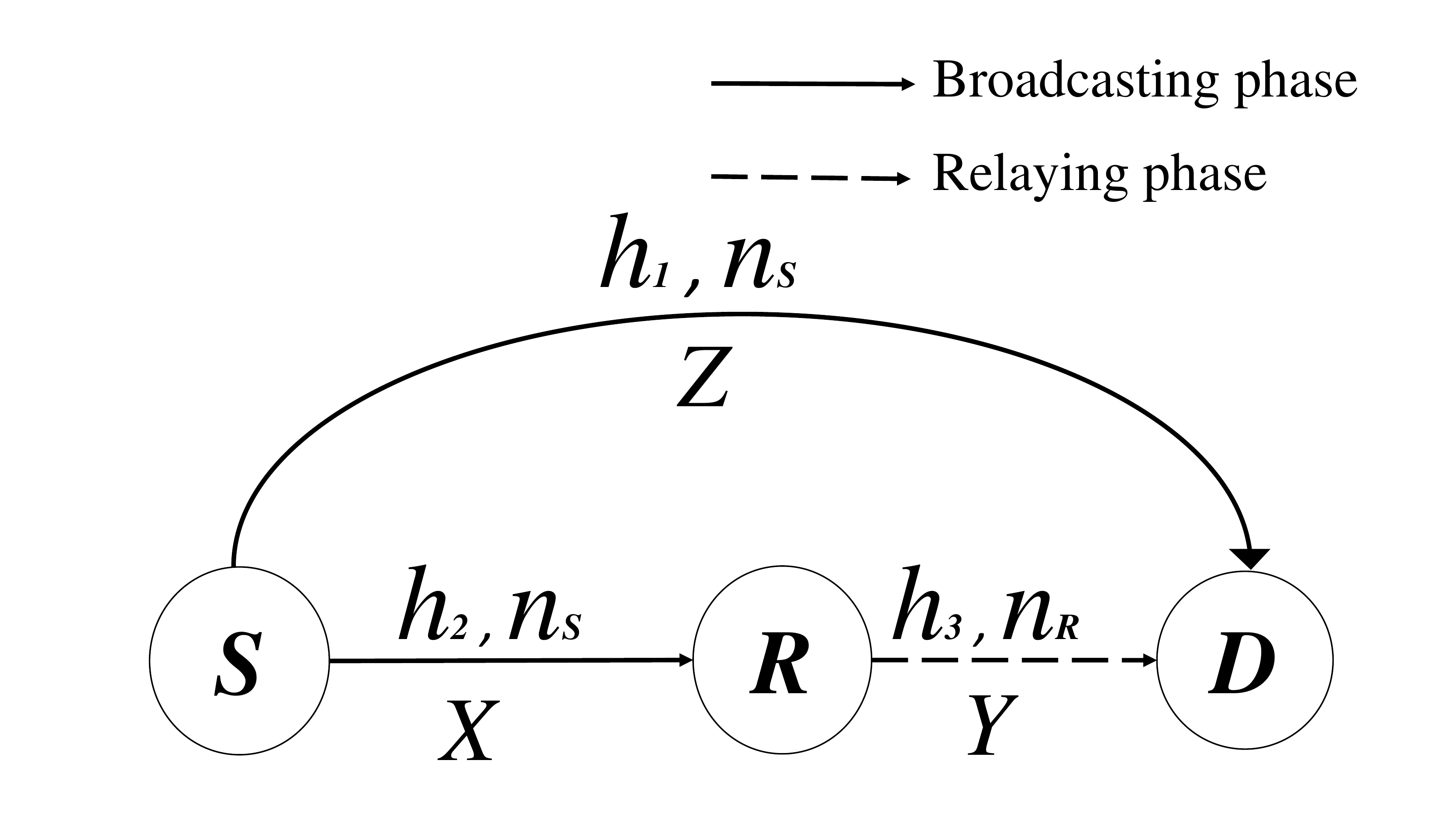}
	\caption{System model for the relaying scenario with a source ($S$), destination ($D$) and a decode-forward relay ($R$). The links between $S$ to $D$, $S$ to $R$ and $R$ to $D$ are referred as the direct link $h_{1}$, broadcasting link $h_{2}$ and relaying link $h_{3}$ each of which with $n_{i}$ channel uses respectively, where ${i \in \{S, R\}}$.}
	\label{fig:System Model}
\end{figure}

Fig.~\ref{fig:System Model} represents a DF relaying scenario that consists of a source $S$, a destination $D$ and a decode-forward relay $R$. The links between $S$-$D$, $S$-$R$ and $R$-$D$ are specified by $Z$, $X$ and $Y$ respectively. In this scenario, $S$ sends the message to $D$ and $R$ in the broadcasting phase by employing $n_{S}$ channel uses. Afterwards, if $R$  decodes the data packet accurately, the message transfers to $D$ in the relaying phase through $n_{R}$ channel uses~\cite{hu2015performance},~\cite{7569613}.
In the broadcasting phase, the received signals at $D$ and $R$ are $y_{1} \!=\! h_{1}x \!+\! w_{1}$ and $y_{2}\!=\! h_{2}x \!+\! w_{2}$ respectively. If $R$ successfully collaborates with $S$, the received signal at $D$ is $y_{3}\!=\! h_{3}x \!+\! w_{3}$, where the transmitted signal is $x$ and the AWGN noise vector with power $N_0\!=\!1$ is denoted by $w_i$, where $i\in \{X,Y,Z\}$~\cite{hu2015performance},~\cite{7569613}.
In a DF-based relaying protocol, the instantaneous SNR depends on the total power constraint $P\!=\!P_{S} \!+\! P_{R} \!=\! \eta P\!+\! (1-\eta) P$, which is given by $Z\!=\! \eta P|h_{1}|^2/N_{0}$, $X\!=\! \eta P|h_{2}|^2/N_{0}$ and $Y\!=\! (1-\eta) P|h_{3}|^2/N_{0}$ for the direct link, broadcasting link and relaying link respectively, where $0<\eta \leq 1$ is the power allocation factor. Thus, the average SNRs are respectively,  $\Omega_Z \!=\! \eta P /N_0$, $\Omega_X \!=\!\eta P /N_0$, $\Omega_Y \!= \!(1\!-\!\eta) P /N_0$. Furthermore,we determine the $S$-$R$ and $R$-$D$ distances according to the normalized $S$-$D$ distance $d_{SD}\!=\!1$, thus $d_{SR}\!=\!\beta$ and $d_{RD}\!=\!1-\beta$ respectively. 
\subsection{Performance Analysis of Single-Hop Communication under the Finite Blocklength Regime} \label{sc:Finite Block length}
In a single-hop communication, $k$ information bits $\mathbf{B}\!=\!\{B_1,..., B_k\}$ pass over an encoder to create a codeword $\mathbf{T}\!=\!\{T_1,..., T_n\}$ with length of $n$ symbols to send over the wireless channel. Subsequently, the channel outputs $\mathbf{D}\!=\! \{D_1,..., D_k\}$ traverse a decoder in order to generate the estimate of the information bits  $\hat{\mathbf{B}}\!=\!\{\hat{B_1},..., \hat{B_k}\}$~\cite{7529226}. Thus, for a single-hop communication with blocklength $n$, outage probability $\epsilon$ and the average power constraint $\rho$, where $\frac{1}{n}\sum_{i}^{n}|x_{i}|^2\leq\rho$ holds, the maximum coding rate  ${\cal R}^*(n,\epsilon)$ of the AWGN channel in bits is calculated as
\begin{equation} \label{eq:maximum rate}
{\cal R}^*(n,\epsilon)\!=\! C(\rho) - \sqrt{\frac {V(\rho)}{n}}\operatorname{Q}^{-1}(\epsilon)\log_{2}\operatorname{e},
\end{equation} 
where, $C(\rho)\!=\!\log_{2}(1+\rho)$ is the positive channel capacity and $V(\rho)\!=\!\rho (2+\rho) \big/(1+\rho)^2$ is the channel dispersion~\cite{7529226}.
According to (\ref{eq:maximum rate}), the outage probability is given by 
\begin{align} \label{eq:outage probability}
\epsilon \!=\! \operatorname{Q}\Bigg(\sqrt{n} \frac{C(\rho)-{\cal R}^*(n,\epsilon)}{\sqrt{V(\rho)}\log_{2}\operatorname{e}}\Bigg),
\end{align}
which holds for the AWGN channels where the channel coefficient $h_{i}$ is equal to one. While for fading channels, we attain the error probability as follow\footnote{This approximation is accurate for n > 100, as proved for AWGN channels \cite[Figs. 12 and 13]{polyanskiy2010channel}, as well as for fading channels as discussed in~\cite{yang2014quasi}.}
\begin{equation} \label{outage_fading}
\epsilon\approx \operatorname{E}\Bigg[\operatorname{Q}\Bigg(\sqrt{n}\frac{C(\rho|h|^2)-{\cal R}^{*}(n,\epsilon) }{\sqrt{V(\rho|h|^2)}}\Bigg)\Bigg].
\end{equation}

Note that in the relaying schemes, we assume that $S$ can encode $k$ information bits into $n_{S}$ channel uses, while $R$ uses $n_{R}$ channel uses. Hence, $S$ and $R$ could employ more sophisticated encoding technique than~\cite{hu2015performance},~\cite{hu2016blocklength}. 
\subsection{Closed-Form Expression of the Outage Probability} \label{Q-function} 
Unfortunately, (\ref{outage_fading}) does not have a closed-form expression; therefore, we resort to an approximation  of the Q-function as in~\cite{7106474},~\cite{6888474}. To do so, let us define  $g(x)= \sqrt{n}\tfrac{C(\rho)-\cal R}{\sqrt{V(\rho)}}$, then we attain
\begin{align}\label{eq:W(t)}
K(t) \approx \operatorname{Q}(g(x))\!=\!
\left\{
\begin{array}{@{}ll@{}}
\ 1 & t\leqslant\varrho\\
\ \dfrac{1}{2}-\dfrac{\mu}{\sqrt{2\pi}}(x-\theta) & \varrho<t< \vartheta\\
\ 0 & t\geq \vartheta
\end{array}\right.
\end{align}
where, $\theta\!=\!\tfrac{2^{\cal R}-1}{P}$, $\vartheta\!=\!\theta\!+\!\sqrt{\tfrac{\pi}{2}\mu^{-2}}$, $\varrho=\theta\!-\!\sqrt{\tfrac{\pi}{2}\mu^{-2}}$ and $\mu =\sqrt{\tfrac{n}{2\pi}}(\operatorname{e}^{2\cal R}-1)^{-\frac{1}{2}}$.
Hence, the outage probability in (\ref{outage_fading}) is defined for Rayleigh fading channels as
\begin{align}\label{outage_fading_linear}
\epsilon&\!=\! \operatorname{E_{X}}[\operatorname{Q}\left(g(x)\right)]\!=\! \int_{0}^{\infty}K(t)f_{X}(x)dx\nonumber \\
& =1\!-\!\dfrac{\zeta}{\sqrt{2\pi}}\exp(-\theta)\bigg[\exp\bigg(\sqrt{\dfrac{\pi}{2\zeta^2}}\bigg)\!-\!\exp\bigg(-\sqrt{\dfrac{\pi}{2\zeta^2}}\bigg)\bigg],
\end{align}
where $f_{X}(x)$ is the PDF of the SNR of the link $X$, $\theta$ is defined in (\ref{eq:W(t)}) and $\zeta= P\sqrt{2\pi}\mu$, where $\mu$ is specified in (\ref{eq:W(t)}). The proof can be found in~\cite{6888474}.

\begin{figure*}[!t]
	\centering
	\vspace{1mm}
	\begin{equation}\label{eq:Outage_MRC_overall}
	\epsilon_{SRD}\!=\! 
	\left\{
	\begin{array}{@{}ll@{}}
	\ 1\!-\!\dfrac{\operatorname{e}^{-\tfrac{\varrho}{\Omega_{Z}}}\!+\! \operatorname{e}^{-\tfrac{\vartheta}{\Omega_{Z}}}}{2}\!-\!\dfrac{\varrho \operatorname{e}^{-\tfrac{\varrho}{\Omega_{Z}}}}{\Omega_{Z}} \!+\!\dfrac{\varrho \operatorname{e}^{-\tfrac{\varrho}{\Omega_{Z}}}-\vartheta \operatorname{e}^{-\tfrac{\vartheta}{\Omega_{Z}}}}{2\Omega_{Z}}\!+\! \dfrac{\mu\theta\left(\operatorname{e}^{-\tfrac{\varrho}{\Omega_{Z}}}\!-\!\operatorname{e}^{-\tfrac{\vartheta}{\Omega_{Z}}}\right)}{\sqrt{2\pi}}\!+\!\dfrac{2\mu\xi}{\sqrt{2\pi}}\!+\!\dfrac{\mu\tau}{\Omega_{Z}\sqrt{2\pi}}&\Omega_{Z}\!=\!\Omega_{Y}\\
	\\
	\
	\dfrac{1}{\Omega_{Z}\!-\!\Omega_{Y}}\bigg[\Omega_{Z}\!-\!\Omega_{Y}\!+\!\Omega_{Z}\operatorname{e}^{-\tfrac{\vartheta}{\Omega_{Z}}}\lambda_{1}\!+\!\Omega_{Z}\operatorname{e}^{-\tfrac{\varrho}{\Omega_{Z}}}\lambda_{2}\!+\!\Omega_{Y}\operatorname{e}^{-\tfrac{\vartheta}{\Omega_{Y}}}\lambda_{3}\!+\!\Omega_{Y}\operatorname{e}^{-\tfrac{\varrho}{\Omega_{Y}}}\lambda_{4}\bigg]&\Omega_{Z}\!\neq\!\Omega_{Y}\	
	\end{array}\right.
	\end{equation}
	\hrule
\end{figure*}
\section{Performance Analysis of the Cooperative Relaying} \label{sec:Relaying assumptions}
In this section, we investigate the outage probability of DF, SC and MRC protocols under the FB regime. The direct transmission model is used here as the basis of the comparison analysis done in this paper. 
\subsection{Dual Hop Decode-and-Forward (DF)}
In this scheme, since the $S$-$D$ distance is too large, it assumes that the direct link is in the outage; thus, $R$ always collaborates with the source. Hence, $S$ sends the message to both $R$ and $D$ in the broadcasting phase. Then, $R$ transfers the message to $D$~\cite{1362898}. The overall outage probability is given by
\begin{equation} \label{DF outage}
\epsilon_{DF}\!=\!\epsilon_{SR}+(1-\epsilon_{SR})\epsilon_{RD},	
\end{equation}
where $\epsilon_{SR}$ and $\epsilon_{RD}$ are calculated according to (\ref{outage_fading_linear}). Notice that we update $\zeta$ with $P_{S}\!=\!\eta P$, $P_{R}= (1\!-\!\eta) P$ and $\mu$ with $n\!=\!n_{S}$, $n\!=\!n_{R}$ respectively. This scenario can be analyzed as selection combining (SC) or maximum ratio combining (MRC) depending on how the destination combines the original transmitted signal and the retransmitted signal.
\subsection{Selection Combining (SC)}
In this protocol, $R$ starts to collaborate with $S$ when the direct transmission is unsuccessful. Therefore, if the $R$-$D$ transmission also fails, $S$ forwards the subsequent frame~\cite{5956530},~\cite{alves2012throughput}. The overall outage is given by\\	
\begin{equation}\label{eq:SDF_outage}
\epsilon_{SC}\!=\!\epsilon_{SD}\epsilon_{SR}+(1-\epsilon_{SR})\epsilon_{SD}\epsilon_{RD},
\end{equation} 
where $\epsilon_{SD}$ is equal to (\ref{outage_fading_linear}) where $\zeta$ is updated with $P_{S}\!=\!\eta P$ and $\mu$ with $n\!=\!n_{S}$.
\subsection{Maximum Ratio Combining (MRC)}\label{sec:MRC}
In this scenario, we have the combined channel gains of $S$-$D$ and $R$-$D$ links at the receiver. Hence, the instantaneous SNR is $\Omega_{W}\!=\! \Omega_{Z}\!+\!\Omega_{Y}$~\cite{5956530},~\cite{alves2012throughput}.
The  outage probability is~\cite{alves2012throughput}
\begin{equation}\label{eq:MRC_outage}
\epsilon_{MRC} \!=\!\epsilon_{SD}\left(\epsilon_{SR}\!+\!\left(1\!-\!\epsilon_{SR}\right)\frac{\epsilon_{SRD}}{\epsilon_{SD}}\right),
\end{equation}
where $\epsilon_{SRD}$ is the outage probability of the source-to-relay-to destination link, notice that the term $\tfrac{\epsilon_{SRD}}{\epsilon_{DF}}$ refers to the probability that D was not able to decode S message alone.
In order to calculate the (\ref{eq:MRC_outage}), first we need to attain the PDF of $W$, which is provided in lemma~\ref{lemma1} and then we calculate the outage probability as proposed in proposition~\ref{propose1}. 
\begin{lemma}\label{lemma1}
Let $W$ denote the sum of two independently distributed exponential RVs, $Z$ and $Y$. Then, $f_W (w)$ is~\cite{alves2012throughput} 	
\begin{equation}\label{eq:MRC_PDF}
	f_W{(w)} \!=\!
	\left\{
	\begin{array}{@{}ll@{}}
	\ \dfrac{w}{\Omega_{Z}^2} \operatorname{exp}\left(-\dfrac{w}{\Omega_{Z}}\right)&\Omega_{Z}\!=\!\Omega_{Y}\\
	\ \dfrac{\operatorname{exp}\left( -\dfrac{w}{\Omega_{Z}}\right)\!-\!\operatorname{exp}\left(- \dfrac{w}{\Omega_{Y}}\right) }{\Omega_{Z}\!-\!\Omega_{Y}} &\Omega_{Z}\!\neq\! \Omega_{Y}\\
	\end{array}\right.
\end{equation}
\begin{proof}
Since the RVs are independent, the proof is straightforward solution of 
	$f_W(w)\!=\! \int_0^\infty  f_Z(w-y) f_y(y) \mathsf{d} y$~\cite{athanasios2017probability}. 
\end{proof}	
\begin{proposition}\label{propose1}
The outage probability of the MRC of the $S$-$D$ and $R$-$D$ links $\epsilon_{SRD}$, is equal to (\ref{eq:Outage_MRC_overall}) on the top of this page, where
\begin{equation}
\tau\!=\!\vartheta^{2}\operatorname{e}^{-\tfrac{\vartheta}{\Omega_{Z}}}-\varrho^{2}\operatorname{e}^{-\tfrac{\varrho}{\Omega_{Z}}}-\theta\vartheta \operatorname{e}^{-\tfrac{\vartheta}{\Omega_{Z}}}+\theta\varrho \operatorname{e}^{-\tfrac{\varrho}{\Omega_{Z}}},
\end{equation}
\begin{equation}
\xi\!=\!\vartheta \operatorname{e}^{-\tfrac{\vartheta}{\Omega_{Z}}}-\varrho \operatorname{e}^{-\tfrac{\varrho}{\Omega_{Z}}}+\Omega_{Z} \operatorname{e}^{-\tfrac{\vartheta}{\Omega_{Z}}}-\Omega_{Z} \operatorname{e}^{-\tfrac{\varrho}{\Omega_{Z}}}, 
\end{equation}	
where, $\vartheta$ and $\varrho$ are specified in (\ref{eq:W(t)}), and $\lambda_{1}\!=\!\tfrac{\mu\vartheta+\mu \Omega_{Z}-\mu\theta}{\sqrt{2\pi}}\!-\!\tfrac{1}{2}$, $\lambda_{2}\!=\!\tfrac{\mu\theta-\mu\varrho-\mu \Omega_{Z}}{\sqrt{2\pi}}\!-\!\tfrac{1}{2}$, $\lambda_{3}\!=\!\tfrac{1}{2}\!-\!\tfrac{\mu\vartheta+\mu \Omega_{Y}-\mu\theta}{\sqrt{2\pi}}$ and $\lambda_{4}\!=\!\tfrac{1}{2}\!+\!\tfrac{\mu\varrho+\mu \Omega_{Y}-\mu\theta}{\sqrt{2\pi}}$.
\end{proposition}	
\begin{proof}
By plugging (\ref{eq:MRC_PDF}) into (\ref{outage_fading_linear}) and multiplying by the linearized Q-function $K(t)$, we attain
\begin{equation}
\epsilon\!=\!\int_{0}^{\varrho}f_X{(x)} dx+ \int_{\varrho}^{\vartheta} \left(\dfrac{1}{2}- \frac{\mu}{\sqrt{(2\pi)}}\left(x-\theta\right)\right)f_X{(x)}dx,
\end{equation}
which is solved with help of~\cite[Eq. 2.321]{gradshteyn2014table} 
and after some algebraic manipulations we attain (\ref{eq:Outage_MRC_overall}).	
\end{proof}
\end{lemma}
\subsection{Direct Transmission}
The source sends the message directly to the destination, where $Z'\!=\!Z/\eta=P|h_{SD}|^2/N_0$, with average SNR $\Omega_Z'\!=\!P/N_0$, where the outage probability is calculated as in (\ref{outage_fading_linear}) but $\zeta$ with $P_{S}\!=\!P$ and $\mu$ with $n\!=\!n_{S}$. 
\begin{figure*}[!t]
	\vspace{3mm}
 	\includegraphics[width=2\columnwidth,height=3.9in]{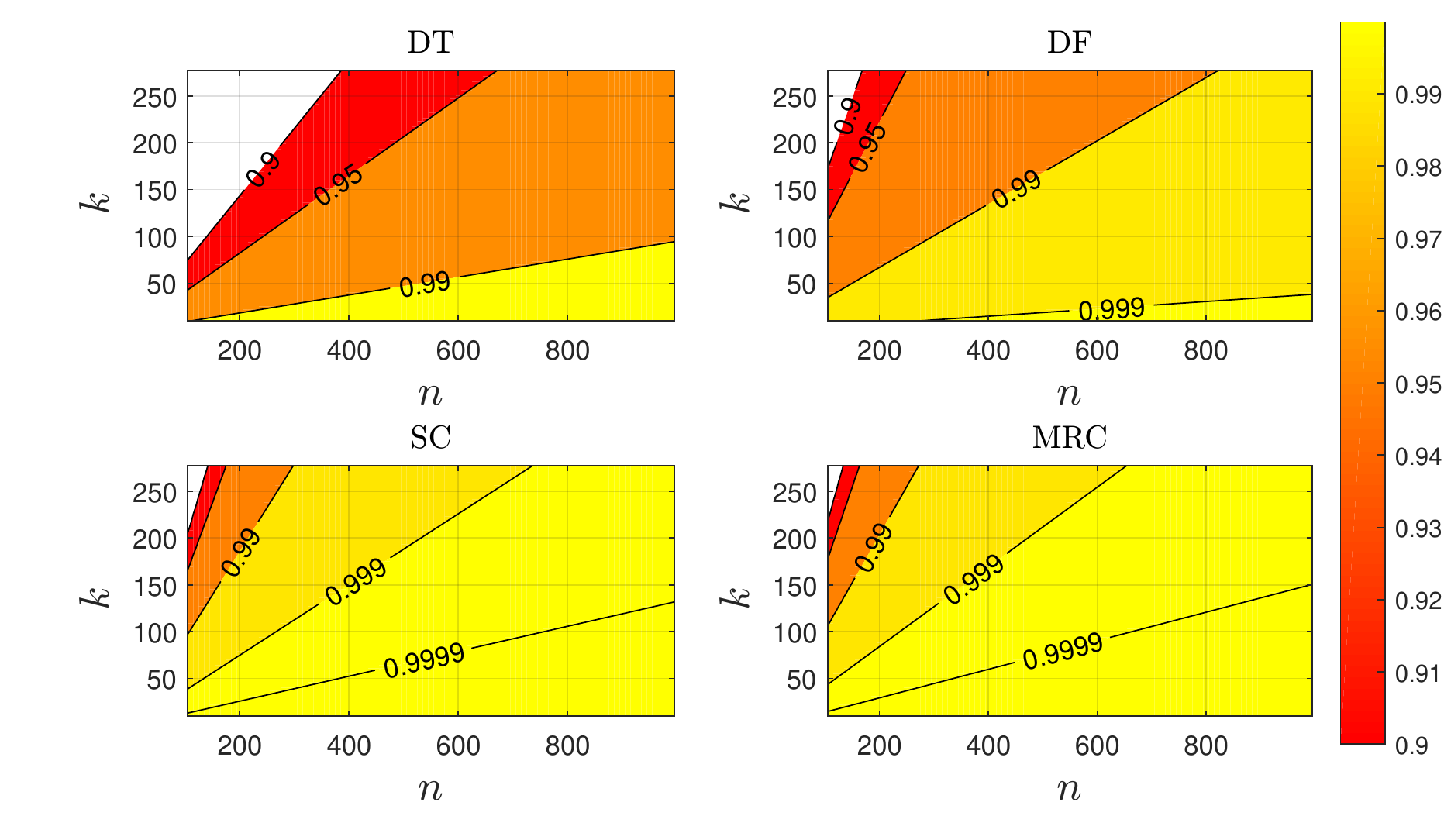}
  	\vspace{-2mm}
 	\caption{Probability of successful transmission in DT, DF, SC and MRC protocols as a function of $k$ and $n$ in the UR region with average SNR as $10$ dB.}\label{fig:contour}
\end{figure*}
\section{Numerical Results}\label{sc:result}
In this paper, we study the performance of the cooperative transmissions under the FB regime for the quasi-static fading channels. First we show that the reliability issue is considerably ameliorated via MRC scenario. Then we examine the impact of the power allocation factor $\eta$ between $S$ and $R$ on the error probability. Thereafter,the performance of the relaying scenarios is compared to the direct transmission in the blocklength Further, we verify via Monte Carlo simulations that the analytical and numerical results properly match. Unless stated; otherwise, assume average SNR of 10 dB, $n\!=\!500$, $\cal R$ $\!=\!0.5$, and $R$ is in between $S$ and $D$, with $\beta\!=\!1/2$.
\subsection{Reliability Improvement via Relaying }
Fig.\ref{fig:contour} compares the impact of the coding rate on the performance of the relaying scenarios under the UR region. It shows the superiority of the MRC protocol over SC, DF and DT scenarios particularly with short data packets under the FB regime. For instance, with packet length $n=200$, $99.99 \%$  reliable transmission is feasible with $k\!=\!31$ and $k\!=\!25$ via MRC and SC protocols respectively, while the probability of the reliable transmission decreases respectively to $99 \%$  with $k\!=\!67$ and $k\!=\!19$ for DF and DT scenarios. Therefore, the performance loss of the MRC is lower with the coding rate growth in comparison to the SC and DF protocols. Moreover, it can be clearly seen that the cooperative protocols have improved the reliability in comparison to the DT.
\begin{figure}[!t]
	 \centering
	\includegraphics[width=1.05\columnwidth]{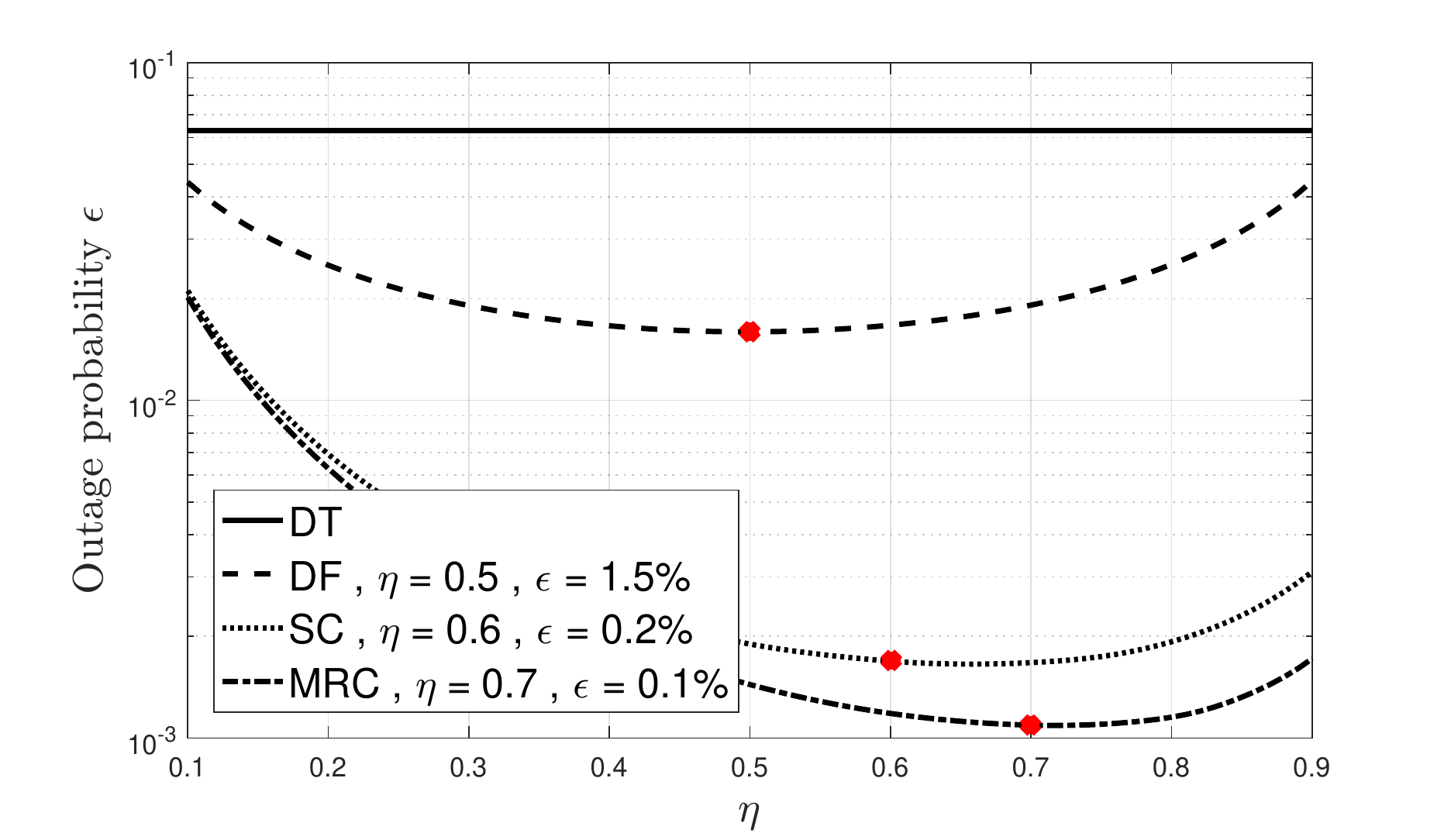}
	\caption{Outage probability variations against the power allocation factor $\eta$ for the average SNR as $10$ dB and $n\!=\!500$ channel uses and $\cal R$ $\!=\!0.5$ bpcu or $k=250$ information bits. The points marked in red are the minimum outage probabilities for the DF, SC and MRC schemes.}\label{fig:betaANDeta_outage}
\end{figure}
\subsection{Power Allocation Factor}
In the cooperative transmissions, we consider a power allocation factor in order to provide a fair comparison between the cooperative protocols and the direct transmission. In Fig.\ref{fig:betaANDeta_outage}, we examine the impact of $\eta$ on the outage probability which is minimized to $1.5
 \%$ (thus $98.5 \%$ reliability) when equal power allocation is employed only for the DF scheme. In SC and MRC, we exploit the additional diversity of the direct link; therefore, more power should be allocated to $S$ as shown in Fig.\ref{fig:betaANDeta_outage}; thus, it would be hard to achieve URC without relaying. For instance, the outage probability is minimized to $0.1 \%$ and $0.2 \%$ with $\eta\!=\! 0.7$ and $\eta\!=\!0.6$ for the MRC and SC schemes respectively. Here, we assume average SNR of $10$ dB and $n\!=\!500$ and $\cal R$$\!=\!0.5$. Similar conclusion holds for other values of SNR, $n$ and $\cal R$. 
\begin{figure}[!t]
	\centering
	 	\vspace{2mm}
	\includegraphics[clip,width=1.05\columnwidth]{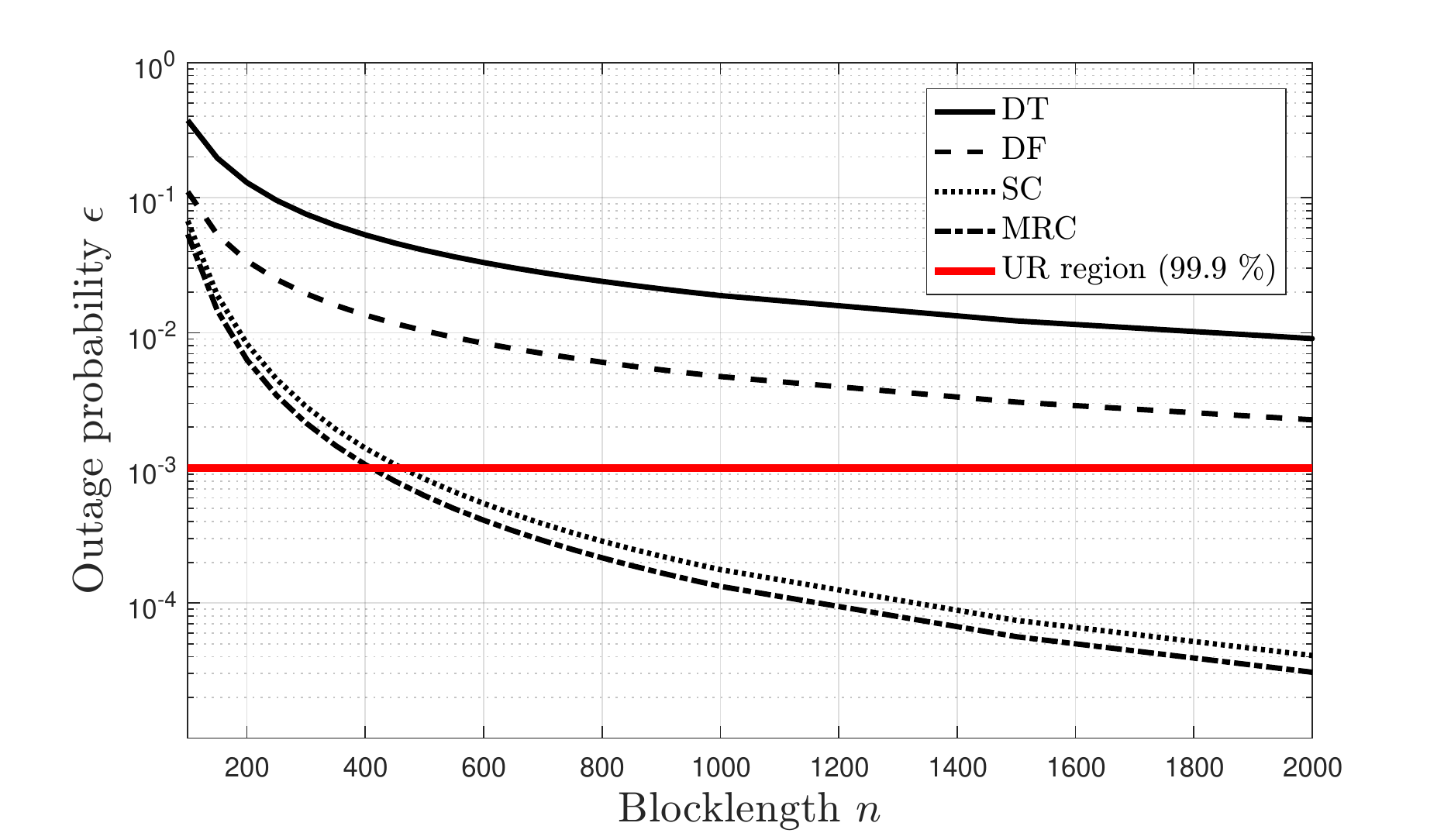}
	\caption{Outage probabilities for DT, DF, SC and MRC scenarios for SNR $\!=\!10$ dB and $k\!=\!250$.}\label{fig:Fading}	
\end{figure}
\subsection{On the Impact of Finite Blocklength}
Here, we define the URC region as reliability > $99.9\%$ and that border line is illustrated in red in the next figures. Fig.\ref{fig:Fading} indicates the performance advantage of the cooperative transmissions over the direct transmission under the FB regime. Outage probability is highly decreased due to the additional spatial diversity. More importantly, the URC is feasible via MRC and SC protocols. 

\begin{figure}[!t]
 	\centering
 	\vspace{2mm}
 	\includegraphics[clip,width=1.05\columnwidth]{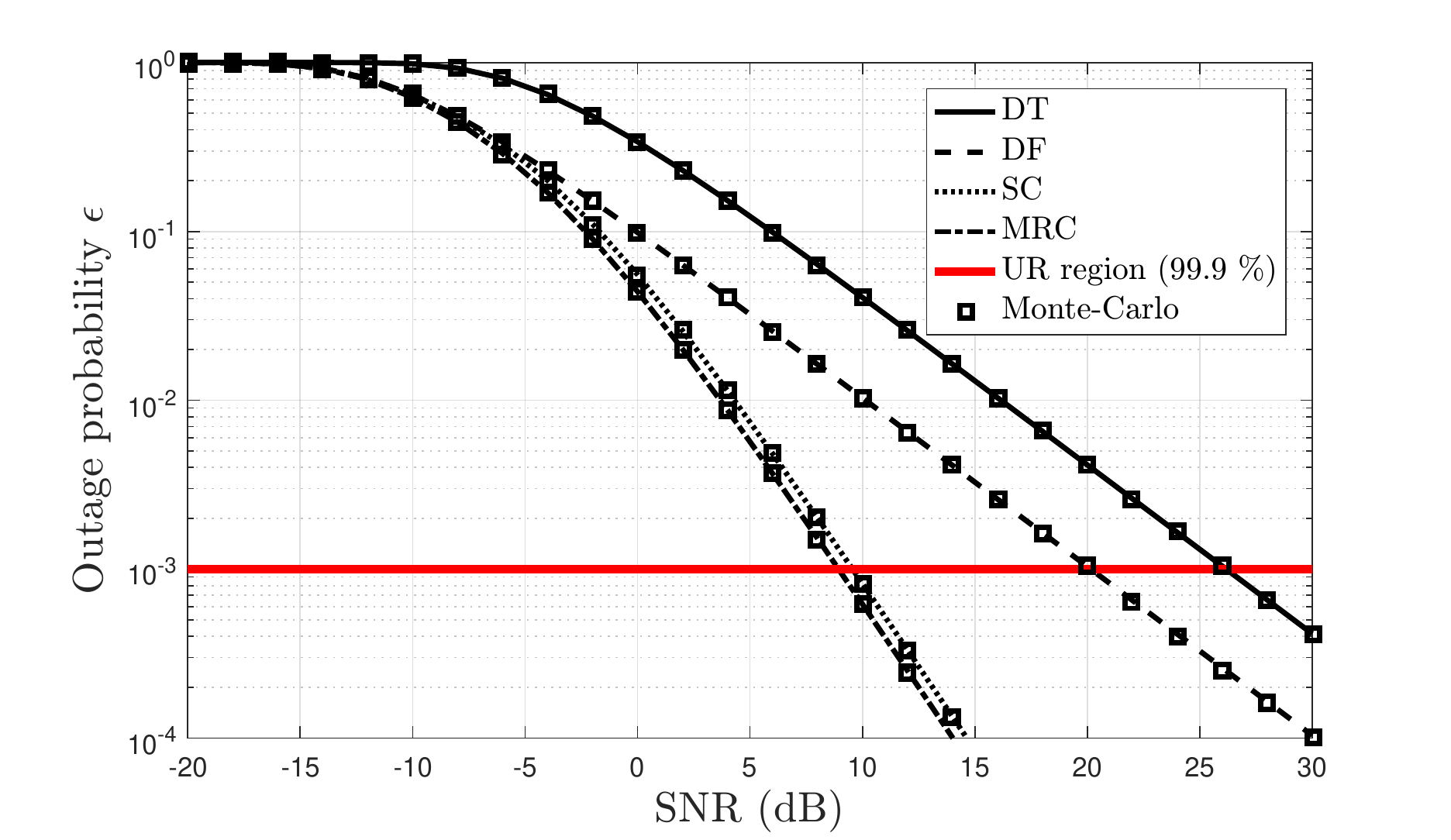}
 	\caption{Comparison between the analytical and numerical results of the outage probabilities for $n\!=\!500$ and $k\!=\!250$.}\label{fig:montecarlo&approx}
\end{figure}

Fig.\ref{fig:montecarlo&approx}, demonstrates a noticeable performance gap among the relaying scenarios and DT, particularly at high SNR regime. Moreover, MRC protocol performs better than the SC and DF in terms of power consumption in the UR region. More importantly, the numerical and analytical results are properly matched due to the Q-function linearization.
\section{Conclusions and Final Remarks} \label{sc:conc}
Herein, we assess the relay communication under the finite blocklength regime. The performance of the different relaying schemes, namely, DF, SC and MRC are compared to the direct transmission in terms of the overall outage probability for the quasi-static Rayleigh fading channels. Moreover, we provide the closed form expression of the overall outage probability. In addition, through the numerical results, we indicate the accuracy of our analytical model and the appropriateness of the outage probability analysis. Notably, we show  that the relaying is a desirable technique in order to improve the system quality through warranting the ultra-high reliability. The superiority of the MRC over the SC, DF and DT is more evident when the packet length is short under the FB regime. In addition, relaying requires less transmit power to communicate in the ultra-high reliable region. Finally, in our future work, we will consider a general case in the context of FB with the imperfect CSI assumption, we will account for the overheads imposed by the relay protocol, in order to provide a desirable trade-off between the reliability and latency requirements in the ongoing and future wireless systems.

\section*{Acknowledgments}
This work has been partially supported by Finnish Funding Agency for Technology and Innovation (Tekes), Huawei Technologies, Nokia and Anite Telecoms, and Academy of Finland (under Grant no. 307492).
\bibliographystyle{IEEEtran}
\bibliographystyle{IEEEtran}

\end{document}